\documentclass[aps,amsmath,amssymb,prl,twocolumn,preprintnumbers,superscriptaddress,nobalancelastpage]{revtex4}
\usepackage{graphicx}% Include figure files
\newcommand{\be}{\begin{eqnarray}}
\newcommand{\ee}{\end{eqnarray}}

\begin{document}
\title{
 Systematic Series Expansions for Processes on Networks
}
\author{
M.~B.~Hastings
       }
\affiliation{
Center for Nonlinear Studies and Theoretical Division, Los Alamos National
Laboratory, Los Alamos, NM 87545, hastings@cnls.lanl.gov
            }
\date{October 21, 2005}
\begin{abstract}
We use series expansions to study dynamics of equilibrium
and non-equilibrium systems on networks.  This analytical method
enables us to include detailed non-universal effects of the
network structure.  We show 
that even low order calculations produce results which compare accurately
to numerical simulation, while the results can be systematically improved.
We show that certain commonly accepted analytical results for the critical point
on networks with a broad degree distribution need to be modified in certain
cases due to disassortativity;
the present method is able to take into account the assortativity
at sufficiently high order, while previous results correspond to
leading and second order approximations in this method.  Finally, we apply this
method to real-world data.
\end{abstract}
\maketitle

Historically, high temperature series expansions played an important
role in the development of the theory of critical phenomena on
regular lattices\cite{domb}, showing non-classical
critical behavior and giving the first hints at
the concept of universality.   The success of these methods is remarkable
when one considers that accurate results near the critical temperature
require taking into account critical fluctuations on all scales.  Such
results were obtained by the ratio method, Pade approximants\cite{baker},
and other means of extrapolating a series of finite length.

In this Letter, we apply series expansions to systems on networks.  One
expects the results to be more accurate at a given order on network systems
than on regular lattices because in many cases these critical correlations
are absent.  On scale-free lattices there are no critical correlations
due to the absence of a spatial structure\cite{sfree}.
On a small-world network\cite{sw}, there are correlations
determined by the regular lattice up to a certain scale and then mean-field
correlations beyond.  In the limit of a small density of long-range links on the
small-world lattice, the correlations on the regular lattice lead to
an anomalous dependence of the mean-field amplitudes on the
link density\cite{mft}.  For a large density of links, the mean-field behavior
takes effect beyond a short scale, and thus we expect even a few terms
of the high temperature series to suffice for accuracy.

Further, the series expansion method includes non-universal effects
including details of the lattice structure.  We will mostly focus on locating
the critical point where we see that the series expansion method
provides quantitatively accurate results that can be systematically improved
by going to higher orders.  Historically, series expansion methods
have provided some of the most accurate results for critical
exponents\cite{domb}, so on this much simpler problem we expect that high
temperature series can be more accurate that direct Monte Carlo simulation.

We begin by developing the series expansions for equilibrium problems,
focusing on an Ising model as a definite example, and then develop the
series expansion for non-equilibrium problems.  We will see that many
of the commonly accepted results for the critical point appear in the
leading orders of the series expansion, while higher orders can significantly
change the location of the critical point due to disassortativity in certain
networks\cite{dis}.

We are aware of one previous use of series for network problems\cite{hts},
for bond percolation, equivalent to certain models of epidemic disease.
In this case, an exact relation was developed between the susceptibility of
a regular lattice and the critical point of a small world network, and the
susceptibility was determined using series expansion.  We show that for
a class of problems which we term ``factorizable" similar exact relations
can be found on small world networks.  For example, we exactly
locate the critical point in the Ising model on a one-dimensional
small-world network.  The final result in this Letter is to show that the
series expansion can be applied directly to an arbitrary real-world
network with accurate results.

{\it Equilibrium Problems---}
We consider an Ising model on a network as a typical
equilibrium problem.  We set all bonds to the same strength for simplicity
to study the effect of network structure.  We leave the network
structure completely arbitrary here, and later give the results in
certain special cases of the network structure.

The partition function is
\be
Z=\sum_{\{\sigma_i\}} \exp[J/kT \sum_{<i,j>} \sigma_i \sigma_j],
\ee
where the sum is over nearest neighbors $i,j$ on the given network.
We locate the critical point by calculating a high temperature
series for the susceptibility, defined by
$\chi=(kT)^{-1} N^{-1} \sum_{i,j} \langle \sigma_i \sigma_j \rangle$ where
$N$ is the total number of sites.  The susceptibility series is
well-suited for locating the critical point because it diverges as
$(T-T_c)^{-1}$\cite{sfree} in these models, while the specific heat
shows a jump at $T_c$ but gives no divergence above $T_c$.

Following standard techniques
of high temperature series expansions\cite{wortis}, we calculate 
$\chi$ to fourth order:
\begin{eqnarray}
\label{chiseries}
kT \chi=1+w \overline{n_1} +w^2 \overline{n_2} + \\ \nonumber
w^3 \overline{n_3}+w^4 (\overline{n_4}
-\overline{n_{\bigtriangleup}})+...,
\end{eqnarray}
where $w=\tanh(J/kT)$ and
where the numbers $\overline{n_m}$ characterize the network by giving the
average over all sites $i$ of the number of self-avoiding paths of length $m$
connected to site $i$.  Thus, $\overline{n_1}=\overline{k}\equiv
N^{-1} \sum_i k_i$, where
$k_i$ is the connectivity of a site $i$.  One may see that
$\overline{n_2}=\overline{k^2-k}\equiv
N^{-1} \sum_i (k_i^2-k_i)$.  The number $n_3$ is equal to
$N^{-1} 
\sum_{i,j,k; <i,j>, <j,k>, i\neq k} 1$, the sum over triplets
$i,j,k$ such that $i\neq k$ and $i,j$ and $j,k$ are pairs of nearest neighbors.
We define
$\overline{n_{\bigtriangleup}}=
N^{-1} \sum_{i,j,k; <i,j>, <j,k>, <i,k>} 1$, the average number of
length-$3$ closed loops.
Thus, for a plane triangular lattice, $n_{\bigtriangleup}=12$.

The simplest method of extracting the critical temperature is the
ratio method: given a series $\sum_n a_n w^n$ given
by Eq.~(\ref{chiseries}), we approximate the
radius of convergence, and hence the critical $w$, by
by $a_n/a_{n-1}$.  The first ratio, $a_1/a_0=\overline{k}$, giving
$w^{-1}=\overline{k}$, and for $\overline{k}$ large this
reduces to $kT_c\approx J\overline{k}$.  The second ratio, $a_2/a_1$, gives
$w^{-1}=\overline{k^2-k}/\overline{k}$.  This second result is the
commonly claimed result\cite{sfree} for the critical point, improving on the
naive result $\overline{k}$.  We will see, however, that even higher orders
in the ratios significantly modify this result by including effects
of assortativity.

A better method of extracting the critical temperature is by means of
Pade approximants\cite{baker}.  We write a series with a finite
number of terms as
$\sum_{n=0}^{L+M} a_n w^n=[\sum_{i=0}^L p_i w^i]/
[\sum_{j=0}^M q_j w^j]+{\cal O}w^{L+M+1}$.  The above series with
four terms can be approximated by an $L=2$, $M=2$ Pade approximant.  The
closest singularity is at
$w^{-1}=
\frac{-a_2a_3+a_1a_4+\sqrt{(a_2a_3-a_1a_4)^2-4(a_2a_4-a_3^2)(a_1a_3-a_2^2)}}
{2(a_2a_4-a_3^2)}$.

{\it Illustration---}
We now apply these results to a simple class of networks with a broad,
but finite, degree distribution and with nontrivial assortativity.
We designate each node with probability
$p$ as a ``high degree" node, and with probability $1-p$ as a ``low
degree" node.  For each high degree node, we assign $m$ connections from
that node to other, randomly chosen, nodes.  If two nodes become connected
more than once via this procedure, we give them only a single connection.
For $N$ large, a low degree nodes has a Poisson distribution of links,
with average degree $pm$ and a high degree node has $k$ links where
$k-m$ is chosen from a Poisson distribution with average $pm$.
Thus, $\overline{n_1}=2pm$, $\overline{n_2}=\overline{k^2-k}=
3 p^2m^2+p(m^2-m)$, $\overline{n_3}= 4p^3m^3+4p^2(m^2-m)(m)$,
and $\overline{n_4}= 5p^4m^4+10p^3(m^2-m)m^2+p^2(m^2-m)^2$.

For $p$ small and $m$ large the degree distribution becomes
very broad and
$\overline{n_1}\approx 2pm, \overline{n_2}\approx pm^2,
\overline{n_3}=4p^2m^3,\overline{n_4}=p^2m^4$.  Thus, the ratio
method for the critical point yields a series of approximations:
$a_1/a_0\approx 2pm, a_2/a_1\approx m/2, a_3/a_2\approx 4pm,
a_4/a_3\approx m/4$, and so on, with an obvious alternation from
term to term.
Using the Pade approximant, we find in the limit $pm>>1, p<<1$ that
\be
\label{rms}
kT_c \approx J \sqrt{\overline{k^2}}.
\ee 

This result (\ref{rms}) for $T_c$
is far below the second order ratio result, $kT_c \approx J
\overline{k^2-k}/\overline{k}$
for the given limit.  The physical reason for this difference is the
disassortativity of the given network.  The high degree nodes connect
to other nodes chosen at random, which are most likely to be low
degree nodes.  In the given limit, most of the paths of length
$2k$ consist of a path that alternates between high and low degree nodes,
and thus there are of order $p^k m^{2k}$ such paths, while there are
only $p^k m^{2k-1}$ paths of length $2k-1$.

Following \cite{dis}, we measure the assortativity of the network via
the connected correlation function of remaining degrees at the end
of a random edge.
This quantity is equal to
\be
(\overline{n_3}+\overline{n_{\bigtriangleup}})/\overline{n_1}-
(\overline{n_2}/\overline{n_1})^2.
\ee
Thus, on a network with vanishing assortativity and no length
three loops, the ratios $a_3/a_2$
and $a_2/a_1$ are equal.  If the higher order ratios,
$a_{n}/a_{n-1}$, are also equal then the critical $w^{-1}$ is correctly
given by $\overline{n_2}/\overline{n_1}$.

In fact, the Pade approximant method is capable of yielding highly
accurate numbers for specific systems even at this low order if the
degree distribution is broad enough.  We choose the specific case
of $p=.1,m=100$ so that $\overline{n_1}=20,\overline{n_2}=1290,\overline{n_3}=
43600,\overline{n_4}=2020100$.  The ratio method gives the series
of different ratios for the critical $w^{-1}$: $20, 64.5, 33.8, 46.33$,
which show alternation from term to term but appear to be converging.
The Pade approximant puts the critical temperature, $T_c=41.46$.
We show in Fig.~1 curves for the specific heat and averge of the
absolute value of the magnetization for
a Monte Carlo simulation of $16384$ spins on the given network.
It can be seen that the Pade result is very close to the correct critical
temperature, while the first and second ratio results $kT_c/J\approx 19.98$ 
and $kT_c/J\approx 64.49$ are poor estimates.

\begin{figure}[tb]
\centerline{
\includegraphics[scale=0.3,angle=0]{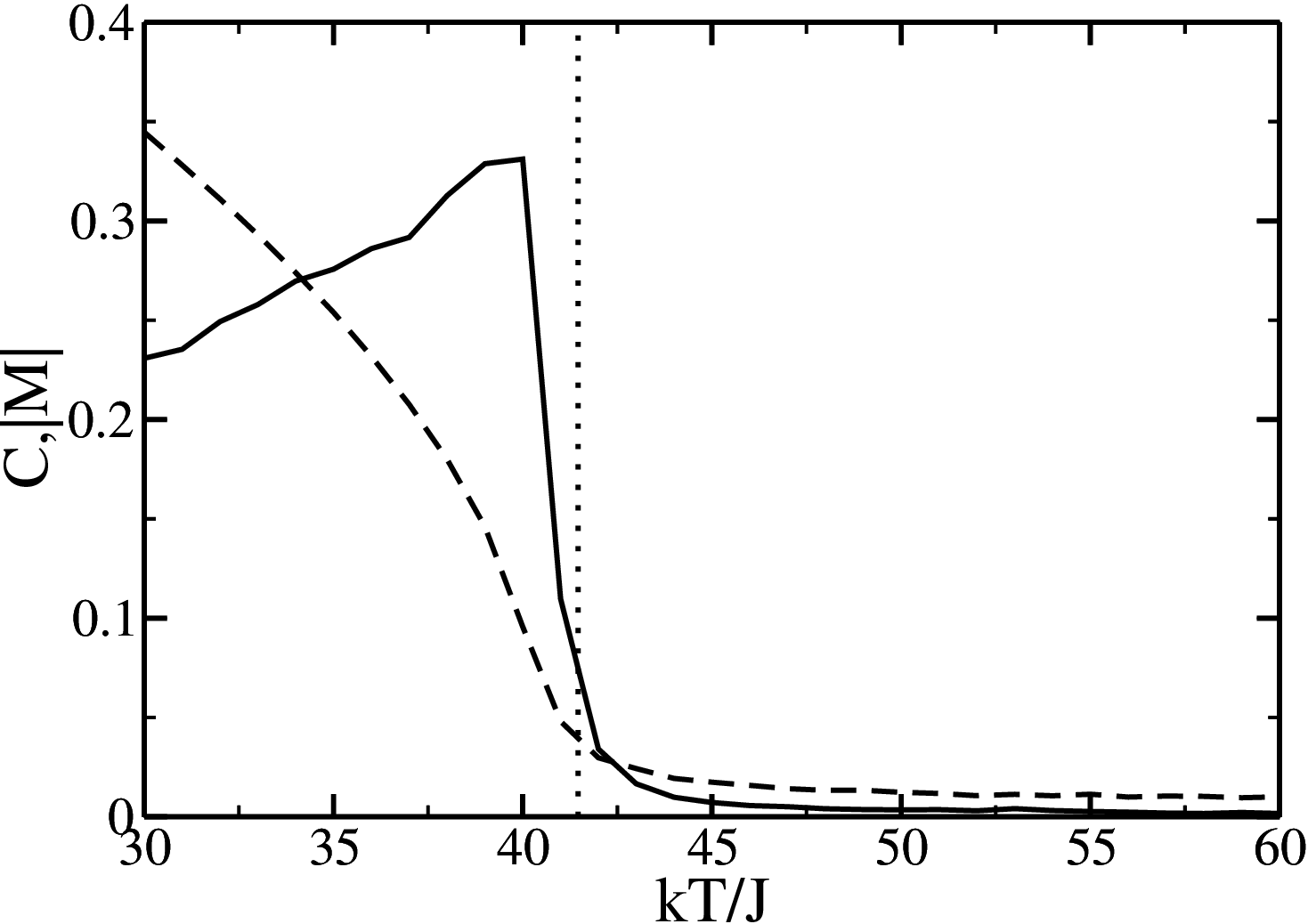}
}
\caption{Specific heat (solid line) and average of absolute
value of magnetization (dashed line) for Monte Carlo simulation of
$16384$ spins.  Pade approximant predicts critical point at
$kT_c/J=41.46$, shown as vertical dotted line.
}
\label{Fig1}
\end{figure}

{\it Exact Results in High Temperature Regime---}
In some cases, the high temperature series can be used to exactly
locate the critical point.  
For the specific
network discussed above, in the limit of large $N$ the loops in the
network become very long and thus looplessness is a good approximation.
One can show exactly that in this model $\overline{n_k}\sim
(pm+\sqrt{p(m^2-m})^k$.
In the Ising model, if the network is loopless,
then the susceptibility series becomes $kT\chi=\sum_k w^k \overline{n_k}$,
and thus the critical point is located at 
$w{-1}\sim\lim_{k\rightarrow\infty} \overline{n_k}^{1/k}$.  
In fact, this result precisely agrees with
the Pade result above, so that the critical point located above is exact.

A second class of problems for which one can find exact results for the
critical point consists of certain ``factorizable" models
on small-world networks\cite{sw} constructed according to the following
procedure: one starts with a regular lattice in $d$-dimensions, and then
adds long-range links, connecting any given pair of nodes with probability
$p/N$.  Consider the high temperature series expansion of
an Ising model on such a network.  
At any fixed, finite order of the expansion for $kT \chi$,
the only loops which appear in the limit of large $N$,
are due to loops on the local lattice\cite{mn}, with no loops due to
shortcuts.  Then, the ensemble average of the susceptibility in
the high temperature regime on the small-world network in the large $N$ limit
is exactly given by
\be
\label{sus}
kT \chi=kT\chi_0(T)/[1-p \tanh(J/kT) kT\chi_0(T)],
\ee
where $\chi_0(T)$ is the susceptibility at temperature $T$
on the local lattice with no shortcuts added.
This is the same kind of result found in \cite{hts}, and such results
can be found for percolation, Ising, and Potts models, as well as the SIR
model, but not the contact process or Heisenberg model where
the series expansions can involve multiple
crossing of the same small-world link.  Similarly, models such as
diffusion\cite{diffusion} on a small-world network
again do not have such an exact result because the model of diffusion
in the absence of long-range links does not have a meaningful series expansion
due to the presence of an infinite correlation length.

Eq.~(\ref{sus}) accords with the mean-field theory result\cite{mft}.
It locates the critical temperature of the one-dimensional small-world
Ising model exactly: $p \tanh(J/kT) \exp[2 J/kT]=1$, so that for
small $p$, $kT=2 J /\log(p^{-1})+...$.

{\it Non-Equilibrium Problems---}
We now consider non-equilibrium processes on networks.  The power series for
these processes can be calculated just as for equilibrium processes, although the
combinatorics are slightly more difficult.  We consider the contact process\cite{cp}
and SIR model\cite{gb}.
In the contact process,
sites are labeled either susceptible or infected, and an infected site can infect
susceptible neighbors at rate $\lambda$, while it can recover and become
susceptible
with unit rate.  In the SIR model, the infected sites become removed, a third
state.  Once
a site reaches the removed state, that node can no longer become infected or
infect its neighbors.

In the contact process, we start at time $t=0$
in a state with all sites susceptible and one
site infected and then compute
\be
\chi_{SIS}\equiv\int_0^{\infty} {\rm d}t \, \sum_i n_i(t),
\ee
where $n_i$ is the probability that site $i$ is infected at time $t$, averaged
over all possible different initially infected sites.
This quantity $\chi_{SIS}$ diverges at the epidemic threshold.
One finds to second order
\be
\label{csis}
\chi_{SIS}=1+\lambda \overline{n_1} + \lambda^2 \overline{n_2} + ...
\ee

For the SIR model, at $t=\infty$ all nodes are either susceptible or removed with
no nodes susceptible.  We define in this case $\chi_{SIR}$ to be the average number
of removed nodes at $t=\infty$ averaged over the same set of initial state, giving
\be
\label{csir}
\chi_{SIR}=1+\lambda \overline{n_1} + \lambda^2 (\overline{n_2}-\overline{n_1}) + ...
\ee

Finally, we consider bond percolation, which is related to, but not identical to,
the SIR model, and in fact has a different series expansion.  Bonds are active with
probability $p$ and we define $\chi_{perc}$ to be the average number of sites connected to a
given starting site.  Here
\begin{eqnarray}
\label{cbp}
\chi_{perc}=1+p\overline{n_1}+p^2\overline{n_2}+
p^3 (\overline{n_3}-\overline{n_{\bigtriangleup}})+\\ \nonumber
p^4 (\overline{n_4}-(3/2)
\overline{n_{\Box}}-2 \overline{n_{\bigtriangleup,1}})+...,
\end{eqnarray}
where
$\overline{n_{\Box}}=
N^{-1} \sum_{i,j,k,l; <i,j>, <j,k>, <k,l>, <l,i>, i\neq k, j\neq l} 1$,
the average number of
length-$4$ closed loops, and
where
$\overline{n_{\bigtriangleup,1}}=
N^{-1} \sum_{i,j,k,l; <i,j>, <j,k>, <k,l>, <l,j>, i\neq k, i\neq l} 1$.
Thus, for a plane square lattice, $n_{\Box}=8$.

At higher orders, the detailed series for these three different processes show
differences, but the dominant term will be of the form $\lambda^k (\overline{n_k}+...)$, so
that if the network is highly connected with $n_{k}>>n_{k-1}$,
we can locate the critical point at $\lambda\approx \lim_{k\rightarrow
\infty} (n_{k-1}/n_{k})$.  Thus, the same results as in the Ising model case for the effects
of assortativity occur again here.

{\it Epidemics on Real-World Networks---}
We now consider the dynamics on real-world networks.
We consider bond percolation as a version of the SIR process on the
283 node connected component of the
web of romantic relationships at ``Jefferson High"\cite{jh}.  Shown in
Fig.~2 is the results of simulation as well as the Pade approximant results.
We include the $[0,1]$ Pade approximant,
$\chi_{perc}\approx 1/(1-p\overline{n_1})$, as well as the
$[1,1]$ approximant, $\chi_{perc}\approx 
[1+(\overline{n_1}-\overline{n_2}/\overline{n_1})p]/
[1-(\overline{n_2}/\overline{n_1})p]$, as well as the $[2,2]$ approximant
to the fourth order series (\ref{cbp}).  The $[0,1]$ and $[1,1]$ results give
the standard results for the critical point, while the $[2,2]$ approximant
is seen to be much more accurate.

\begin{figure}[tb]
\centerline{
\includegraphics[scale=0.3,angle=0]{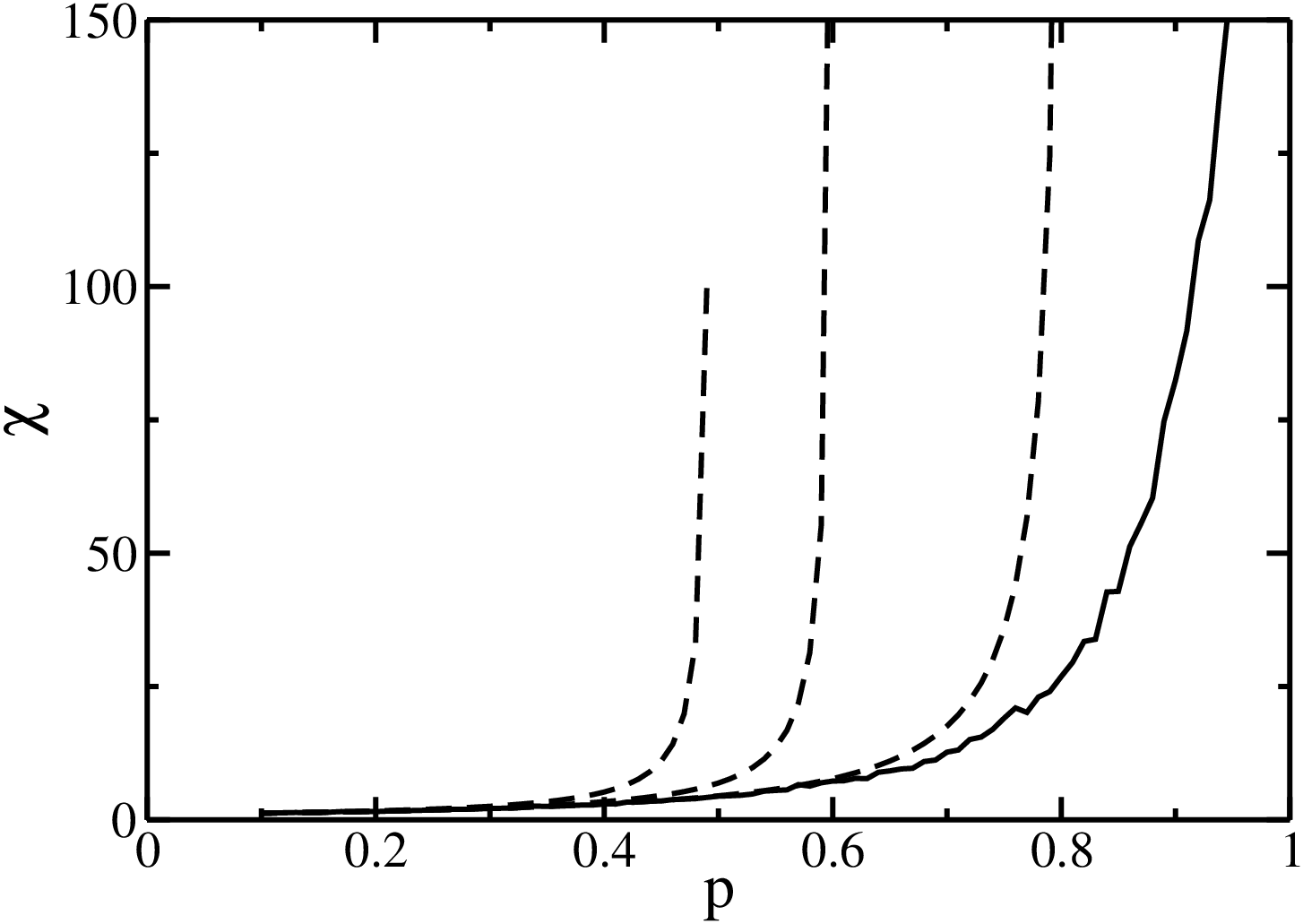}
}
\caption{Susceptibility of bond percolation on social network.  Solid line is
from simulation.  Dashed lines are (from left to right) $[0,1],[1,1],[2,2]$
Pade approximant results.}
\label{Fig2}
\end{figure}

We believe that the difference between the $[2,2]$ approximant and
the simulation is in fact due to finite size effects.  Differences set
in only when a sizable fraction of the network is on average part of the
percolation cluster.  We believe that if larger social networks were
obtained, of the order of the number of people in a typical city, the $[2,2]$
approximant would become highly accurate.
However, it is difficult
to get large data on accurate social networks to test the series technique
against simulation.  In fact, this is precisely the reason for using the
series technique: it allows one to obtain accurate results from {\it local}
data.

Thus, to illustrate finite size effects, we return to the network
defined in ``Illustration", taking $p=.2$ and $m=10$.  We find the results
shown in Fig~3.  The $[2,2]$ approximant is again found to be more accurate,
and it is seen that the simulation of $10000$ sites is closer to the Pade
result than the simulation of $1000$ sites.  As before, in fact, the Pade
predicts the exact result for the critical point on this network in the infinite
size limit.

The Pade result above can be determined for any network such that we can
find the various constant $n_1,n_2,...$.  This can be
determined based
on local information and Monte Carlo sampling of a small portion of
the network.

\begin{figure}[tb]
\centerline{
\includegraphics[scale=0.3,angle=0]{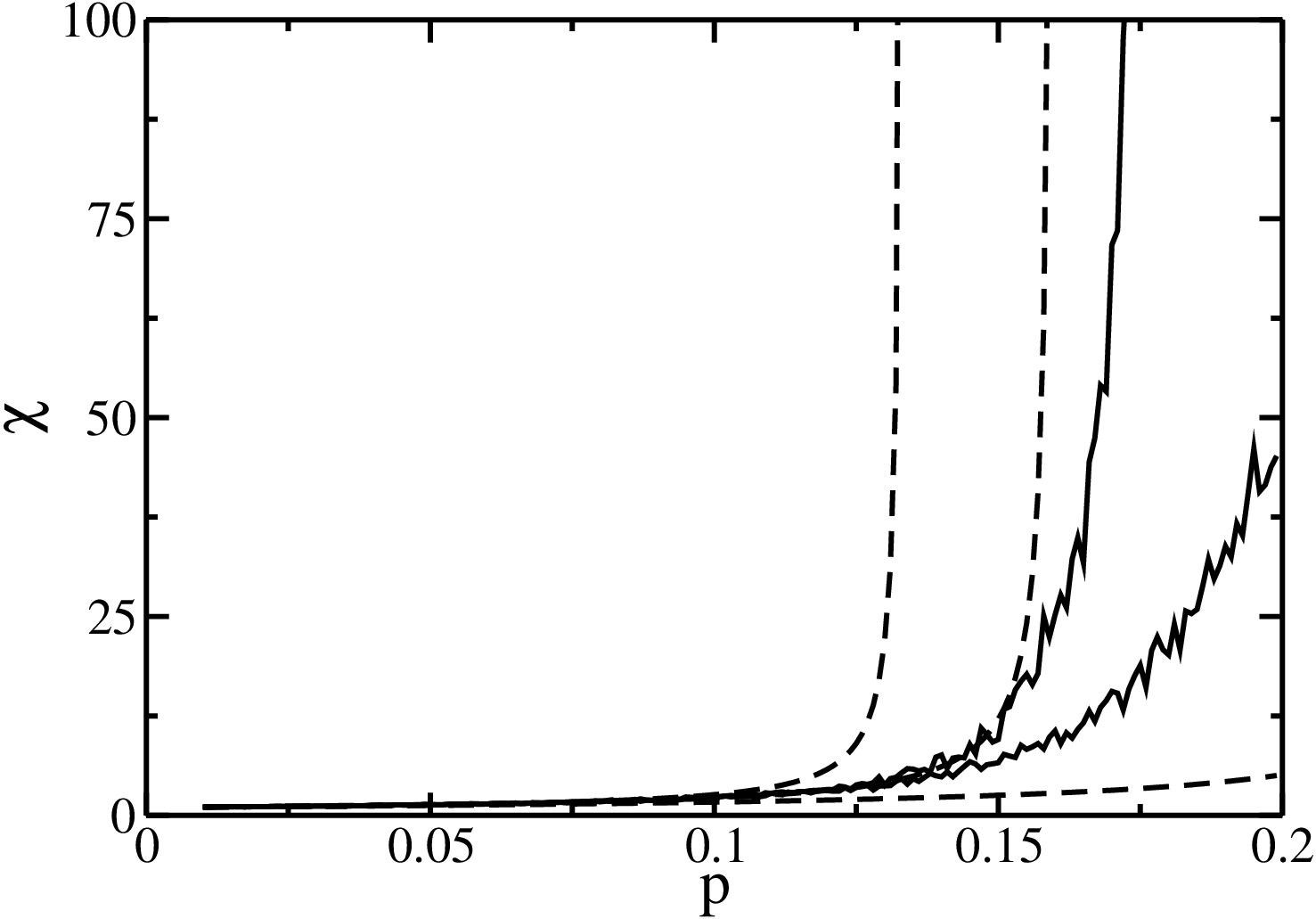}
}
\caption{Susceptibility of bond percolation on simulated network.  Solid lines
are (from left to right)
from simulation of $10000$ and $1000$ nodes.  
Dashed lines are (from left to right) $[1,1],[2,2],[0,1]$
Pade approximant results.}
\label{Fig3}
\end{figure}

{\it Discussion---}
The series expansion technique provides a systematic method for
studying networks in the ``high-temperature" phase.  It is possible
to extend this technique to include the ordered phase.  For example,
in an Ising model one can compute the partition function as
$Z=\langle \exp[-J/kT \sum_{<i,j>} \sigma_i \sigma_j-\sum_i h_i\sigma_i]\rangle_h=
\exp[\sum_n (1/n!) 
\langle (-J/kT \sum_{<i,j>} \sigma_i \sigma_j-\sum_i h_i\sigma_i)^n\rangle_{h,c}]$,
where the expectation value $\langle...\rangle_h$ is with respect to
$\exp[+\sum_i h_i \sigma_i]$, and $\langle...\rangle_{h,c}$ denote various
cumulants.  
Keeping only the first cumulant, setting
$h_i=h(k_i)$ for some function $h$ depending only on the
local coordination number, and maximizing $Z$ over $h$
reproduces previous results\cite{sfree}, while corrections are found at
higher order.

The technique provides exact results on certain systems, such as loopless
networks and certain problems on small-world networks, where it agrees
with, but goes beyond, previous mean-field calculations.  It works well
on simulated and realistic networks.  In many cases, only local information
on a network is available.  A tendency has been then to construct a model
which agrees with the local information, such as distribution of connectivities,
and then simulate the desired process on such a simulated network.  Series
expansions instead directly express the desired result in terms of the local
properties.  One key result is the quantitative measure of how the assortativity
affects the critical properties.

{\it Acknowledgements---}
I thank M. A. Moore for useful discussions on the application of series
expansions to critical phenomena.
This work was supported by DOE grant 
W-7405-ENG-36.  

\end{document}